
!!!!!!!!!!!!!!!!!!!!!!!!!!!!!!!!!!!!!!!!!!!!!!!!!!!!!!!!!!!!!!!!!!!!!!
%
\newif\ifproofmode			
\proofmodefalse				

\newif\ifforwardreference		
\forwardreferencetrue			

\newif\ifchapternumbers			
\chapternumbersfalse			

\newif\ifcontinuousnumbering		
\continuousnumberingfalse		

\newif\iffigurechapternumbers		
\figurechapternumbersfalse		

\newif\ifcontinuousfigurenumbering	
\continuousfigurenumberingfalse		

\newif\ifcontinuousreferencenumbering   
\continuousreferencenumberingtrue       

\newif\ifparenequations			
\parenequationstrue			

\newif\ifstillreading			

\font\eqsixrm=cmr6			
\def\marginstyle{\eqsixrm}		

\newtoks\chapletter			
\newcount\chapno			
\newcount\eqlabelno			
\newcount\figureno			
\newcount\referenceno			
\newcount\minutes			
\newcount\hours				

\newread\labelfile			
\newwrite\labelfileout			
\newwrite\allcrossfile			

\chapno=0
\eqlabelno=0
\figureno=0

%
\def\initialeqmacro{
    \ifproofmode
        \headline{\tenrm \today\ --\ \timeofday\hfill
                         \jobname\ --- draft\hfill\folio}
        \hoffset=-1cm
        \immediate\openout\allcrossfile=zallcrossreferfile
    \fi
    \ifforwardreference
        \openin\labelfile=zlabelfile
        \ifeof\labelfile
        \else
            \stillreadingtrue
            \loop
                \read\labelfile to \nextline
                \ifeof\labelfile
                    \stillreadingfalse
                \else
                    \nextline
                \fi
                \ifstillreading
            \repeat
        \fi
        \ifproofmode
            \immediate\openout\labelfileout=zlabelfile
        \fi
    \fi}


{\catcode`\^^I=9
\catcode`\ =9
\catcode`\^^M=9

%
\gdef\chapfolio{			
    \ifnum \chapno>0 \relax		
        \the\chapno
    \else
        \the\chapletter
    \fi}

%
\gdef\bumpchapno{
    \ifnum \chapno>-1 \relax
        \global \advance \chapno by 1
    \else
        \global \advance \chapno by -1 \setletter\chapno
    \fi
    \ifcontinuousnumbering
    \else
        \global\eqlabelno=0
    \fi
    \ifcontinuousfigurenumbering
    \else
        \global\figureno=0
    \fi
    \ifcontinuousreferencenumbering
    \else
        \global\referenceno=0
    \fi}

%
\gdef\setletter#1{\ifcase-#1 {}  \or\global\chapletter={A}
  \or\global\chapletter={B} \or\global\chapletter={C} \or\global\chapletter={D}
  \or\global\chapletter={E} \or\global\chapletter={F} \or\global\chapletter={G}
  \or\global\chapletter={H} \or\global\chapletter={I} \or\global\chapletter={J}
  \or\global\chapletter={K} \or\global\chapletter={L} \or\global\chapletter={M}
  \or\global\chapletter={N} \or\global\chapletter={O} \or\global\chapletter={P}
  \or\global\chapletter={Q} \or\global\chapletter={R} \or\global\chapletter={S}
  \or\global\chapletter={T} \or\global\chapletter={U} \or\global\chapletter={V}
  \or\global\chapletter={W} \or\global\chapletter={X} \or\global\chapletter={Y}
  \or\global\chapletter={Z}\fi}

%
\gdef\tempsetletter#1{\ifcase-#1 {}\or{} \or\chapletter={A} \or\chapletter={B}
 \or\chapletter={C} \or\chapletter={D} \or\chapletter={E}
  \or\chapletter={F} \or\chapletter={G} \or\chapletter={H}
   \or\chapletter={I} \or\chapletter={J} \or\chapletter={K}
    \or\chapletter={L} \or\chapletter={M} \or\chapletter={N}
     \or\chapletter={O} \or\chapletter={P} \or\chapletter={Q}
      \or\chapletter={R} \or\chapletter={S} \or\chapletter={T}
       \or\chapletter={U} \or\chapletter={V} \or\chapletter={W}
        \or\chapletter={X} \or\chapletter={Y} \or\chapletter={Z}\fi}

%
\gdef\chapshow#1{
    \ifnum #1>0 \relax
        #1
    \else
        {\tempsetletter{\number#1}\the\chapletter}
    \fi}

%
\gdef\today{\number\day\space \ifcase\month\or Jan\or Feb\or
        Mar\or Apr\or May\or Jun\or Jul\or Aug\or Sep\or
        Oct\or Nov\or Dec\fi, \space\number\year}

\gdef\timeofday{\minutes=\time    \hours=\time
        \divide \hours by 60
        \multiply \hours by 60
        \advance \minutes by -\hours
        \divide \hours by 60
        \ifnum\the\minutes>9
		\the\hours:\the\minutes
 	\else
  		\the\hours:0\the\minutes
	\fi}


%
%
\gdef\chapnum{\bumpchapno \chapfolio}			

\gdef\chaplabel#1{                                      
    \bumpchapno                                         
    \ifproofmode                                        
        \ifforwardreference                             
            \immediate\write\labelfile{
            \noexpand\expandafter\noexpand\def
            \noexpand\csname CHAPLABEL#1\endcsname{\the\chapno}}
        \fi
    \fi
    \global\expandafter\edef\csname CHAPLABEL#1\endcsname
    {\the\chapno}	 				
    \ifproofmode
        \llap{\hbox{\marginstyle #1\ }}
    \fi
    \chapfolio}

%
%
\gdef\eqnum{                                    
    \global\advance\eqlabelno by 1              
    \eqno(
    \ifchapternumbers
        \chapfolio.
    \fi
    \the\eqlabelno)}

\gdef\eqlabel#1{                                
    \global\advance\eqlabelno by 1              
    \ifproofmode                                
        \ifforwardreference
            \immediate\write\labelfileout{\noexpand\expandafter\noexpand\def
            \noexpand\csname EQLABEL#1\endcsname
            {\the\chapno.\the\eqlabelno?}}
        \fi
    \fi
    \global\expandafter\edef\csname EQLABEL#1\endcsname
    {\the\chapno.\the\eqlabelno?}
    \eqno(
    \ifchapternumbers
        \chapfolio.
    \fi
    \the\eqlabelno)
    \ifproofmode
        \rlap{\hbox{\marginstyle #1}}		
    \fi}

\gdef\eqalignnum{                               
    \global\advance\eqlabelno by 1              
    &(\ifchapternumbers
        \chapfolio.
    \fi
    \the\eqlabelno)}

\gdef\eqalignlabel#1{                   	
    \global\advance\eqlabelno by 1 	        
    \ifproofmode		  		
        \ifforwardreference
            \immediate\write\labelfileout{\noexpand\expandafter\noexpand\def
            \noexpand\csname EQLABEL#1\endcsname
            {\the\chapno.\the\eqlabelno?}}
        \fi
    \fi
    \global\expandafter\edef\csname EQLABEL#1\endcsname
    {\the\chapno.\the\eqlabelno?}
    &(\ifchapternumbers
        \chapfolio.
    \fi
    \the\eqlabelno)
    \ifproofmode
        \rlap{\hbox{\marginstyle #1}}			
    \fi}

\gdef\eqref#1{\ifparenequations(\fi
    \ifundefined{EQLABEL#1}***
        \ifproofmode
            \ifforwardreference)
            \else
                \write16{ ***Undefined\space Equation\space Reference #1*** }
            \fi
        \else
            \write16{ ***Undefined\space Equation\space Reference #1*** }
        \fi
    \else
        \edef\LABxx{\getlabel{EQLABEL#1}}
        \def\LAByy{\expandafter\stripchap\LABxx}
        \ifchapternumbers
            \chapshow{\LAByy}.\expandafter\stripeq\LABxx
        \else
            \ifnum \number\LAByy=\chapno \relax
                \expandafter\stripeq\LABxx
            \else
                \chapshow{\LAByy}.\expandafter\stripeq\LABxx
            \fi
        \fi
        \ifparenequations)\fi
    \fi
    \ifproofmode
        \write\allcrossfile{Equation #1}
    \fi}

%
\gdef\fignum{                                   
    \global\advance\figureno by 1\relax         
    \iffigurechapternumbers
        \chapfolio.
    \fi
    \the\figureno}

\gdef\figlabel#1{				
    \global\advance\figureno by 1\relax 	
    \ifproofmode				
        \ifforwardreference
            \immediate\write\labelfileout{\noexpand\expandafter\noexpand\def
            \noexpand\csname FIGLABEL#1\endcsname
            {\the\chapno.\the\figureno?}}
        \fi
    \fi
    \global\expandafter\edef\csname FIGLABEL#1\endcsname
    {\the\chapno.\the\figureno?}
    \iffigurechapternumbers
        \chapfolio.
    \fi
    \ifproofmode
        \llap{\hbox{\marginstyle #1\ }}\relax
    \fi
    \the\figureno}

\gdef\figref#1{					
    \ifundefined				
        {FIGLABEL#1}!!!!			
        \ifproofmode
            \ifforwardreference
            \else
                \write16{ ***Undefined\space Figure\space Reference #1*** }
            \fi
        \else
            \write16{ ***Undefined\space Figure\space Reference #1*** }
        \fi
    \else
        \edef\LABxx{\getlabel{FIGLABEL#1}}
        \def\LAByy{\expandafter\stripchap\LABxx}
        \iffigurechapternumbers
            \chapshow{\LAByy}.\expandafter\stripeq\LABxx
        \else \ifnum\number\LAByy=\chapno \relax
                \expandafter\stripeq\LABxx
            \else
                \chapshow{\LAByy}.\expandafter\stripeq\LABxx
            \fi
        \fi
        \ifproofmode
            \write\allcrossfile{Figure #1}
        \fi
    \fi}

%
\gdef\refnum{                                      
    \global\advance\referenceno by 1\relax         
    \the\figureno}	                           

\gdef\internalreflabel#1{			
    \global\advance\referenceno by 1\relax 	
    \ifproofmode	   			
        \ifforwardreference
            \immediate\write\labelfileout{\noexpand\expandafter\noexpand\def
            \noexpand\csname REFLABEL#1\endcsname
            {\the\chapno.\the\referenceno?}}
        \fi
    \fi
    \global\expandafter\edef\csname REFLABEL#1\endcsname
    {\the\chapno.\the\figureno?}
    \ifproofmode
        \llap{\hbox{\marginstyle #1\hskip.5cm}}\relax
    \fi
    \the\referenceno}

\gdef\internalrefref#1{				
    \ifundefined				
        {REFLABEL#1}!!!!			
        \ifproofmode
            \ifforwardreference
            \else
                \write16{ ***Undefined\space Footnote\space Reference #1*** }
            \fi
        \else
            \write16{ ***Undefined\space Footnote\space Reference #1*** }
        \fi
    \else
        \edef\LABxx{\getlabel{REFLABEL#1}}
        \def\LAByy{\expandafter\stripchap\LABxx}
        \expandafter\stripeq\LABxx
        \ifproofmode
            \write\allcrossfile{Figure #1}
        \fi
    \fi}

%
\gdef\reflabel#1{\item{\internalreflabel{#1}.}}

%
\gdef\refref#1{\internalrefref{#1}}

\gdef\eq{\ifhmode Eq.~\else Equation~\fi}		
\gdef\eqs{\ifhmode Eqs.~\else Equations~\fi}

%
%

%
\gdef\getlabel#1{\csname#1\endcsname}
\gdef\ifundefined#1{\expandafter\ifx\csname#1\endcsname\relax}
\gdef\stripchap#1.#2?{#1}				
\gdef\stripeq#1.#2?{#2}					
}  

\magnification=\magstep1
\overfullrule = 0pt
\hoffset = 0.25 truein
\hsize=6.0truein
\vsize=9.0truein
\baselineskip=14truept
\chapternumberstrue
\continuousnumberingfalse
\forwardreferencetrue
\initialeqmacro
\def\sh{\mathop{\rm sh}\nolimits}
\def\ch{\mathop{\rm ch}\nolimits}

\def\tr{\mathop{\rm tr}\nolimits}
\newbox\lett
\newdimen\lheight
\newdimen\lwidth
\def\ontop#1#2{\setbox\lett=\hbox{#2}%
   \lheight\ht\lett \multiply\lheight by 12 \divide\lheight by 10\relax%
   \lwidth\wd\lett \multiply\lwidth by 8 \divide\lwidth by 10\relax%
   #2\kern-\lwidth%
       \raise\lheight\hbox{{$\scriptstyle #1$}}\kern.1ex}
\line{\hfill UMTG-163}
\vglue 1.3 truein
\noindent
EXACTLY SOLVED MODELS WITH QUANTUM ALGEBRA SYMMETRY\footnote*{Presented
at the Miami workshop ``Quantum Field Theory, Statistical Mechanics,
Quantum Groups, and Topology,'' 7 - 12 January 1991.}
\vskip 0.56 truein
\hbox to \hsize{\hbox to 1.2 truein{\hfil}Luca Mezincescu and
Rafael I. Nepomechie\hfil}
\vskip 0.06 truein
\hbox to \hsize{\hbox to 1.2 truein{\hfil}Department of Physics\hfil}
\vskip 0.06 truein
\hbox to \hsize{\hbox to 1.2 truein{\hfil}University of Miami, Coral Gables,
Florida 33124\hfil}
\vskip .23 truein
\noindent
ABSTRACT
\vskip 0.2truein

We have constructed and solved various one-dimensional quantum mechanical
models
which have quantum algebra symmetry. Here we summarize this work, and
also present new results on graded models, and on the so-called string
solutions
of the Bethe Ansatz equations for the $A^{(2)}_2$ model.

\vskip 0.4truein

\noindent
\chapnum . INTRODUCTION
\vskip 0.2truein

The concept of symmetry is fundamental for the description of physical
systems. In many cases, such symmetry is codified by a Lie (super) algebra.
A generalization of this structure, the so-called {\it quantum}
Lie (super) algebra, has recently emerged${}^{\refref{s} -
\refref{woronowicz}}$.
Our ultimate goal is to understand how quantum algebra symmetry is implemented
in physical systems, and to explore the consequences of this symmetry.
Such symmetry may eventually prove to be useful for field theory
in 4 spacetime dimensions and for string theory.

To date, quantum algebras have been identified as the common mathematical
structure linking three types of physical systems: topological (Chern-Simons)
field theory in 3 spacetime dimensions${}^{\refref{witten}}$, integrable
lattice models${}^{\refref{pasquier}}$,
and rational conformal field theories${}^{\refref{rcft}}$ and their integrable
perturbations${}^{\refref{pertb}}$.
Over the past two years, we have studied primarily the connection between
integrable lattice models and quantum algebras.
Among the three connections of quantum algebras to physical systems noted
above, this is the most direct. Furthermore, it is within this context
that quantum algebras were first discovered.

In the course of our investigations${}^{\refref{vladimir}-\refref{analytic}}$,
we have constructed and
solved various one-dimensional quantum mechanical models which have quantum
algebra symmetry. Here we summarize this work, and
also present new results on graded models, and on the string solutions
of the Bethe Ansatz equations for the $A^{(2)}_2$ model.
The construction of these models requires two main ingredients: the
$R$ matrix, which can be interpreted as a two-particle scattering amplitude,
and the $K$ matrix, which can be interpreted as the amplitude for a particle
to reflect elastically
from a wall. The integrability of these models comes from demanding
that the scattering be consistent with factorization. In Section 2,
we introduce $R$ matrices via the Zamolodchikov algebra, and summarize some of
their important properties. In Section 3, we introduce $K$ matrices through an
extension of the Zamolodchikov algebra.
In particular, we describe the graded case, which we illustrate
with an example connected to the superalgebra $su(2|1)$.
In Section 4, we construct open chains of $N$ ``spins'' (generators of a
quantum algebra) with certain nearest-neighbor interactions, which are
integrable and which have quantum algebra symmetry.
For these models, the transfer matrix (i.e.,
not just the Hamiltonian) commutes with the generators of a quantum algebra.
We also comment on the solution -- namely, the eigenvalues of the transfer
matrix and the Bethe Ansatz equations -- of these models. In order to
calculate quantities of physical interest, one must first solve the Bethe
Ansatz equations in the $N \rightarrow \infty$ limit. For the $A^{(1)}_1$
case, these so-called string solutions are well known. In Section 5, which
is a result of a collaboration with A.M. Tsvelik,
we investigate string solutions for the $A^{(2)}_2$ model of
Izergin and Korepin. We find new types
of string solutions, but we are not able to formulate a general string
hypothesis. We summarize our results in Section 6.

A more detailed account for the simplest case of $A^{(1)}_1$ can be found
in Ref. \refref{spin1}.

\vskip 0.4truein

\noindent
\chapnum . $R$ MATRICES
\vskip 0.2truein

\noindent
{\bf Yang-Baxter equation}

\vskip 0.2truein

We briefly review here how the (graded) Yang-Baxter equation follows
from the associativity of the (graded) Zamolodchikov algebra.
This algebra is abstracted from
studies of scattering in massive relativistic quantum field
theories in 1+1 dimensions with an infinite number of conservation
laws${}^{\refref{zamolodchikov}}$.

The Zamolodchikov algebra has generators $A_\alpha (u)$, where $u$
is the so-called spectral parameter, and $\alpha = 1 \,, \dots \,, n$.
These generators obey the relations
$$ A_\alpha (u) \ A_\beta (v) = {}_{\alpha \beta}R_{\alpha' \beta'}(u-v)\
A_{\beta'}(v)\ A_{\alpha'} (u) \,. \eqlabel{zamalgebra} $$
The matrix ${}_{\alpha \beta}R_{\alpha' \beta'}(u-v)$, which may be
interpreted as a two-particle scattering amplitude, is called the $R$ matrix.

By setting $u=v$ in the above relation, and by assuming linear independence of
monomials of second degree, we learn that the $R$ matrix is regular,
$$R(0) = {\cal P} \,, \eqnum $$
where ${\cal P}$ is the permutation matrix,
$$ {}_{\alpha \beta} {\cal P}_{\alpha' \beta'} =
\delta_{\alpha \beta'}\ \delta_{\beta \alpha'} \,.  \eqnum $$
Moreover,
by interchanging $A_\alpha (u) \ A_\beta (v)$ twice using \eqref{zamalgebra},
we obtain the unitarity relation
$$R(u)\  {\cal P}\ R(-u)\ {\cal P} = 1 \,. \eqlabel{unitarity} $$

Consider now the monomial of third degree
$A_\alpha (u) \ A_\beta (v) \ A_\gamma (0)$. Associativity of the
Zamolodchikov algebra, as well as the assumption of linear independence of
monomials of third degree, imply that
$$\eqalignno{
& {}_{\alpha \beta}R_{\alpha'' \beta''}(u - v)\
{}_{\alpha'' \gamma}R_{\alpha' \gamma''}(u)\
{}_{\beta'' \gamma''}R_{\beta' \gamma'}(v) = \cr
& \quad {}_{\beta \gamma}R_{\beta'' \gamma''}(v)\
{}_{\alpha \gamma''}R_{\alpha'' \gamma'}(u)\
{}_{\alpha'' \beta''}R_{\alpha' \beta'}(u-v) \,. \eqalignnum \cr} $$
This relation is the well-known Yang-Baxter (or factorization) equation .
Introducing the notation $R_{12} = R \otimes 1$, so that
$${}_{\alpha \beta \gamma} \left( R_{12} \right)_{\alpha' \beta' \gamma'}
= {}_{\alpha \beta}R_{\alpha' \beta'}\ \delta_{\gamma \gamma'} \,, $$
and similarly defining $R_{13}$ and $R_{23}$, the Yang-Baxter equation
can be rewritten in the compact form
$$R_{12}(u-v)\ R_{13}(u)\ R_{23}(v)\ =
  R_{23}(v)\ R_{13}(u)\ R_{12}(u-v) \,. \eqlabel{yb} $$

We remark that the Yang-Baxter equation transforms covariantly under
 ``gauge'' (or ``symmetry-breaking'') transformations${}^{\refref{jimbo},
\refref{sogo/akutsu/abe}, \refref{nonsymmetric}}$ of the $R$ matrix
$$R_{12}(u - v)\ \rightarrow \ontop1B(u)\
\ontop2B(v)\ R_{12}(u - v)\
\ontop1B(-u)\ \ontop2B(-v) \,, \eqlabel{gauge} $$
where $B(u)$ is a diagonal matrix with the properties
$$B(u)\ B(v) = B(u + v) \,, \quad\quad\quad B(0) = 1, \eqnum $$
as well as
$$\ontop1B(u)\ R_{12} (v)\ \ontop1B(-u) =
\ontop2B(-u)\ R_{12} (v)\ \ontop2B(u) \,. \eqlabel{moreprop} $$
Here we have introduced the notation
$$\ontop1B \equiv B \otimes 1 \,, \quad\quad\quad
  \ontop2B \equiv 1 \otimes B \,. \eqnum $$

There is a graded version of the Yang-Baxter equation.
Following Kulish and Sklyanin${}^{\refref{kulish/sklyanin}}$, we introduce a
$Z_2$ grading of the Zamolodchikov algebra, by considering the
generators $A_\alpha$ to be homogeneous elements with parity
$ p(\alpha) \equiv p( A_\alpha )$ equal to either 0 (even) or 1 (odd).
These generators obey the relations
$$ A_\alpha (u)\ A_\beta (v) =
(-)^{p(\alpha) p(\beta)}\ {}_{\alpha \beta}R_{\alpha' \beta'}(u-v)\
A_{\beta'}(v)\ A_{\alpha'} (u)  \,. \eqlabel{gradedzamalgebra} $$
We assume that ${}_{\alpha \beta}R_{\alpha' \beta'}$ are commuting numbers,
and that if ${}_{\alpha \beta}R_{\alpha' \beta'} \ne 0$, then
$p(\alpha) + p(\beta) + p(\alpha') + p(\beta') = 0 \quad mod \ 2$.
Associativity of this graded Zamolodchikov algebra
leads to the graded Yang-Baxter equation${}^{\refref{kulish/sklyanin}}$,
$$\eqalignno{
&(-)^{p(\beta'')\left[ p(\gamma'') - p(\gamma) \right]}\
{}_{\alpha \beta}R_{\alpha'' \beta''}(u - v)\
{}_{\alpha'' \gamma}R_{\alpha' \gamma''}(u)\
{}_{\beta'' \gamma''}R_{\beta' \gamma'}(v) \cr
& \quad = (-)^{p(\beta'')\left[ p(\gamma'') - p(\gamma') \right]}\
{}_{\beta \gamma}R_{\beta'' \gamma''}(v)\
{}_{\alpha \gamma''}R_{\alpha'' \gamma'}(u)\
{}_{\alpha'' \beta''}R_{\alpha' \beta'}(u-v) \,. \eqalignnum \cr} $$

\vskip 0.2truein

\noindent
{\bf Solutions of the Yang-Baxter equation}

\vskip 0.2truein

An $R$ matrix is said to be quasi-classical if it depends on an additional
parameter $\eta$ which plays the role of Planck's constant, so that
$$R(u, \eta)\Big\vert_{\eta=0} = const \ 1 \,. \eqnum $$
There are three known classes of regular quasi-classical
solutions of the Yang-Baxter equation: elliptic,
trigonometric, and rational (corresponding to the three types of functions
of $u$ that appear in $R(u)$ ).

Being interested in quantum algebras, we focus on the trigonometric
solutions. Such solutions are associated${}^{\refref{belavin/drinfeld}}$
with affine Lie algebras
$g^{(k)}$, where $g$ is a simple Lie algebra ($A_n = su(n+1)$,
$B_n = o(2n + 1)$, $C_n = sp(2n)$, $D_n = o(2n)$, etc.) and
$k (=1,2,3)$ is the order of a diagram automorphism $\sigma$ of $g$.
That is, $\sigma^k = 1$.
The cases $k=1$ and $k>1$ are often referred to as ``untwisted'' and
``twisted'', respectively.
For instance, in the case of $A^{(2)}_2$ in the fundamental representation,
the diagram automorphism is given by the complex
conjugation map $\sigma : \lambda^A \rightarrow -\lambda^{A *}$,
where $\lambda^A$ are the eight Gell-Mann matrices.

We shall later make use of the fact that the automorphism $\sigma$
leaves invariant a subalgebra $g_0$ of $g$. (This subalgebra $g_0$ is
in fact the maximal finite-dimensional subalgebra of the affine algebra
$g^{(k)}$.) In the $A^{(2)}_2$ example,
it is clear that $\sigma$ leaves invariant the purely imaginary matrices
$\lambda^2$, $\lambda^5$, $\lambda^7$, which generate an $su(2)$ subalgebra
of $su(3)$. A table listing every simple Lie algebra $g$ which has a
nontrivial diagram
automorphism, along with the corresponding subalgebra $g_0$ which
is left invariant by this automorphism, is given in Ref. \refref{kac}, and is
reproduced in Ref. \refref{twisted}.

The simplest example of an untwisted $R$ matrix is the spin 1/2
$A^{(1)}_1$ matrix
$$
R^{({1\over2}, {1\over2})}(u)  = {1\over \sqrt{| \sh(u+\eta) \sh(-u + \eta) |}}
\left(
\matrix{\sh (u + \eta)&        &        &  \cr
                   & \sh u    & \sh \eta &  \cr
                   & \sh \eta & \sh u    &  \cr
                   &        &        & \sh(u + \eta) \cr} \right) \,.
\eqlabel{symmetric}
$$
In this gauge, the $R$ matrix is ``symmetric''; i.e., it is both $P$ invariant
(${\cal P}_{12}\ R_{12}\ {\cal P}_{12}  = R_{12}$) and
$T$ invariant ($R_{12}^{t_1 t_2} = R_{12}$). The gauge
transformation \eqref{gauge} with $B(u) = diag (e^{u/2} ,e^{-u/2} )$
yields the symmetry-broken $R$ matrix
$$
R^{({1\over2}, {1\over2})}(u)  = {1\over \sqrt{| \sh(u+\eta) \sh(-u + \eta) |}}
\left(
\matrix{\sh (u + \eta)&        &        &  \cr
                   & \sh u    & e^{u}\sh \eta &  \cr
                   & e^{-u}\sh \eta & \sh u    &  \cr
                   &        &        & \sh(u + \eta) \cr} \right) \,.
\eqlabel{nonsymmetric}
$$
which is only $PT$ invariant,
$${\cal P}_{12}\ R_{12}(u)\ {\cal P}_{12} = R_{12}^{t_1 t_2}(u) \,.
\eqlabel{pt} $$
The transposition $t_i$ refers to the $i^{th}$ space.

The $R$ matrices associated with the nonexceptional affine Lie algebras
in the fundamental representation have been given by
Bazhanov${}^{\refref{bazhanov}}$ and Jimbo${}^{\refref{jimbo}}$. (For the
graded case, see Ref. \refref{bazhanov/shadrikov}.)
Although in general these $R$ matrices
do not have either $P$ or $T$ symmetry, they do have $PT$ symmetry \eqref{pt}.
Except for $A^{(1)}_n$ ($n > 1$), these $R$ matrices have crossing symmetry,
$$R_{12}^{t_1}(u)\ \ontop1M \ R_{12}^{t_2}(-u - 2\rho) \ \ontop1M^{\ -1}
= 1 \,, \eqlabel{crossing} $$
where $M$ is a symmetric matrix ($M^t = M$) which can be deduced from
Ref. \refref{bazhanov}.
Moreover, except for $D^{(2)}_{n}$, these $R$ matrices (in the so-called
homogeneous gauge used by Jimbo) satisfy
$$\left[ \check R(u) \,, \check R(v) \right] = 0 \,, \eqlabel{jimboidentity} $$
where
$$\check R(u) \equiv {\cal P} R(u) \,. \eqnum $$

\vskip 0.2truein

\noindent
{\bf Connection with Quantum Algebras}

\vskip 0.2truein

The prototype quantum algebra is $U_q [su(2)]$, with generators
$\vec S = \left\{ S^+ , S^- , S^z \right\}$ which obey
$$ q^{S^z \mp 1}\ S^\pm  = S^\pm \ q^{S^z} \,, \quad\quad\quad
   \left[ S^+ \,, S^- \right] = {q^{2S^z} - q^{-2S^z} \over q - q^{-1}} \,,
\eqlabel{su(2)q} $$
where $q$ is a complex parameter. Given two sets of generators $\vec S_1$,
$\vec S_2$ of this algebra (with $\left[ \vec S_1 \,, \vec S_2 \right] = 0$),
the generators $\vec S$ in the tensor product space are given by
$$q^{S^z} =  q^{S^z_1} \otimes q^{S^z_2} \,, \quad\quad\quad
S^{\pm} =  S^{\pm}_1 \otimes q^{-S^z_2} + q^{S^z_1} \otimes S^{\pm}_2 \,.
\eqlabel{coproduct}
$$
The generalization to $U_q [g]$ for any simple Lie algebra $g$ is discussed
in Refs. \refref{drinfeld} - \refref{faddeev}.

Faddeev, {\it et al.}${}^{\refref{faddeev}}$ emphasize an $R$-matrix
formulation of quantum algebras. Taking $U_q [su(2)]$ again as an example,
define
$$R_\pm = \lim_{u \rightarrow \pm \infty} R(u) \,, \eqlabel{rplusminus} $$
where $R(u)$ is the spin 1/2 $A^{(1)}_1$ $R$ matrix in the nonsymmetric
gauge \eqref{nonsymmetric}; and define the upper, lower triangular matrices
$$T_+ = \left( \matrix{ q^{{S^z} + {1\over 2}} & (q-q^{-1})S^- \cr
                        & q^{{-S^z} + {1\over 2}} \cr } \right) \,,
\quad\quad\quad
T_- = \left( \matrix{ q^{{-S^z} - {1\over 2}}  \cr
                    -(q-q^{-1})S^+  & q^{{S^z} - {1\over 2}} \cr } \right) \,,
\eqlabel{borel} $$
with $q = e^\eta$. The relations
$$ R_\pm \ \ontop1T_\pm \ \ontop2 T_\epsilon =
\ontop2T_\epsilon \ \ontop1T_\pm \ R_\pm  \quad\quad \hbox{with}
\quad \epsilon = \{+ \,, -\}
\eqlabel{rtt} $$
hold if and only if the operators $\vec S$ obey the algebra \eqref{su(2)q}.
Moreover, consider two sets of such matrices $T_{1 \pm}$, $T_{2 \pm}$
constructed from $\vec S_1$, $\vec S_2$ respectively. The coproduct
matrices $T_\pm$ are given by
$$ T_\pm = T_{1 \pm} \otimes T_{2 \pm} \eqnum $$
where the symbol $\otimes$ indicates the tensor product of the algebras and
the usual product of the matrices. They
are expressed in the form \eqref{borel} in terms of the operators $\vec S$
given precisely by the comultiplication rule \eqref{coproduct}.

An important identity is
$$\left[ \check  R(u) \,, U_q [su(2)] \right] = 0  \,, \eqnum  $$
where here by $U_q [su(2)]$ we mean coproducts of the generators.
For the general case of an $R$ matrix of the type $g^{(k)}$, the
corresponding result is${}^{\refref{jimbo}}$
$$\left[ \check  R(u) \,, U_q [g_0] \right] = 0  \,, \eqlabel{namely}  $$
where $g_0$ is the subalgebra of $g$ which is left invariant under
the diagram automorphism of order $k$.
In particular, for both $A^{(1)}_1$ and $A^{(2)}_2$, the matrices $\check R(u)$
commute with $U_q[su(2)]$.

\vskip 0.4truein

\noindent
\chapnum . $K$ MATRICES
\vskip 0.2truein

\noindent
{\bf Reflection-factorization equation}

\vskip 0.2truein

We now extend the Zamolodchikov algebra \eqref{zamalgebra}, by introducing
the additional relation
$$ A_\alpha (u) = {}_\alpha K_{\alpha'}(u)\ A_{\alpha'}(-u) \,.
\eqlabel{zamextension} $$
The $K$ matrix ${}_\alpha K_{\alpha'}(u)$ can be interpreted as the amplitude
for a particle to reflect elastically from a wall${}^{\refref{zamo}}$.

By setting $u=0$, we see that
$$ K(0) = 1 \,. \eqnum $$
Furthermore, using the relation \eqref{zamextension} twice, we obtain the
unitarity relation
$$ K(u)\ K(-u) = 1 \,. \eqnum $$

Consider now the monomial of second degree $A_\alpha (u)\ A_\beta (v)$.
There are two different ways by which one can apply each of the Zamolodchikov
relations \eqref{zamalgebra}, \eqref{zamextension} twice to obtain
an expression proportional to $A_{\alpha'} (-u)\ A_{\beta'} (-v)$. Using
again the assumption of linear independence, we obtain the
relation${}^{\refref{zamo} - \refref{sklyanin},\refref{nonsymmetric}}$
$$ R_{12}(u-v)\ \ontop1K(u)\ {\cal P}_{12}\ R_{12}(u + v)\ {\cal P}_{12}\
\ontop2K(v) =
\ontop2K(v)\  R_{12}(u + v)\ \ontop1K(u)\ {\cal P}_{12}\ R_{12}(u-v)\
{\cal P}_{12} \,, \eqlabel{sklyanin} $$
to which we shall refer as the reflection-factorization equation.
This equation transforms covariantly under the gauge transformation
\eqref{gauge},
provided that the $K$ matrix transforms as follows,
$$K(u) \rightarrow B(u)\ K(u)\ B(u)   \,. \eqlabel{gaugek} $$

By repeating the above
calculation using instead {\it graded} Zamolodchikov
generators (which obey the relation \eqref{gradedzamalgebra}), we obtain the
graded reflection-factorization equation,
$$\eqalignno{
& (-)^{p(\beta'') p(\alpha''')}\
{}_{\alpha \beta}R_{\alpha'' \beta''}(u-v)\ {}_{\alpha''}K_{\alpha'''}(u)\
{}_{\beta'' \alpha'''}R_{\beta''' \alpha'}(u+v)\ {}_{\beta'''}K_{\beta'}(v)\cr
&  = (-)^{p(\beta''') p(\alpha''')}\
{}_{\beta}K_{\beta''}(v)\ {}_{\alpha \beta''}R_{\alpha'' \beta'''}(u+v)\
{}_{\alpha''}K_{\alpha'''}(u)\ {}_{\beta''' \alpha'''}R_{\beta' \alpha'}(u-v)\
\,. \eqalignnum \cr}
$$
Here we have assumed that ${}_{\alpha}K_{\alpha'}$ are commuting numbers,
and that if ${}_{\alpha}K_{\alpha'} \ne 0$, then
$p(\alpha) + p(\alpha') = 0 \quad mod \ 2$.

\vskip 0.2truein

\noindent
{\bf Solutions of the reflection-factorization equation}

\vskip 0.2truein

Given a solution $R(u)$ of the (graded) Yang-Baxter equation, one can solve
the (graded) reflection-factorization relation for the corresponding $K(u)$.

\medskip

\noindent
{\it spin 1/2 $A^{(1)}_1$ :}

For the spin 1/2 $A^{(1)}_1$ $R$ matrix \eqref{symmetric}, there is a
one-parameter family of diagonal $K$ matrices
given by ${}^{\refref{cherednik},\refref{sklyanin}}$
$$K^{({1\over 2})}(u, \xi) = {1\over \sqrt{| \sh(u + \xi) sh(-u + \xi) |}}
\left( \eqalign{& \sh(u + \xi)\cr
                &\quad\quad -\sh(u - \xi)\cr} \right) \,,
\eqlabel{k1} $$
where $\xi$ is an arbitrary parameter.

\medskip

\noindent
{\it spin 1 $A^{(1)}_1$ :}

For the spin 1 $A^{(1)}_1$ matrix $R^{(1,1)}$ given in Refs.
\refref{fateev},\refref{kulish/reshetikhin}, we find${}^{\refref{spin1}}$
$$K^{(1)}(u, \xi) = \rho (u, \xi)
\left( \eqalign{&\sh (u + \xi) \sh (u - \eta + \xi) \cr
&\quad\quad\quad -\sh (u - \xi) \sh (u - \eta + \xi) \cr
&\quad\quad\quad\quad\quad\quad\quad\quad
\sh (u - \xi) \sh (u + \eta - \xi) \cr} \right) \,, \eqlabel{k2} $$
where
$$\rho (u, \xi) = \left[ \sh(u + \xi) \sh(-u + \xi) \sh(u - \eta + \xi)
\sh(-u - \eta + \xi) \right]^{-{1\over2}} \,. $$
Just as there is a fusion procedure${}^{\refref{fusion}}$ by which
$R^{(1,1)}$ may be obtained from $R^{({1\over2}, {1\over2})}$,
there is a similar fusion procedure${}^{\refref{spin1}}$ by which $K^{(1)}$
may be obtained from $K^{({1\over 2})}$ and $R^{({1\over 2}, {1\over 2})}$.

\vfill\eject

\noindent
{\it $A^{(2)}_2$ in fundamental representation:}

The $R$ matrix corresponding to
$A^{(2)}_2$ in the fundamental representation${}^{\refref{izergin},
\refref{kulish/sklyanin}}$ is the simplest example of a ``twisted'' $R$ matrix.
For this case, we obtain${}^{\refref{twisted}}$
$$K(u, \pm) = \rho (u, \pm)
\left( \eqalign{
&\ch ( {u\over 2} + 3\eta) \mp i \sh {u\over 2} \cr
&\quad\quad\quad
e^{-u} \left[ \ch( {u\over 2} - 3\eta) \pm i \sh {u\over 2} \right] \cr
&\quad\quad\quad\quad\quad\quad\quad\quad
e^{u} \left[ \ch( {u\over 2} - 3\eta) \pm i \sh {u\over 2} \right] \cr}
\right) \,, \eqnum $$
where
$$\rho (u, \pm) = \left(
\left[ \ch ( {u\over 2} + 3\eta) \mp i \sh {u\over 2} \right]
\left[ \ch ( {u\over 2} - 3\eta) \pm i \sh {u\over 2} \right]
\right)^{-{1\over2}} \,. $$
In contrast with the $A^{(1)}_1$ case, for which there is a one-parameter
family
of ``nontrivial'' (i.e., $K \ne 1$) diagonal solutions, here we find only two
nontrivial solutions.

\medskip

\noindent
{\it $sl(2|1)^{(2)}$ in fundamental representation:}

For the solution \eqref{gradedrmatrix}
of the graded Yang-Baxter equation corresponding
to $sl(2|1)^{(2)}$ in the fundamental representation, we find the following
one-parameter family of solutions of the graded reflection-factorization
equation:
$$K(u, \xi) = \rho (u, \xi)
\left( \eqalign{&\sh (u + \xi) \ch (u - \eta + \xi) \cr
&\quad\quad\quad -\sh (u - \xi) \ch (u + \eta - \xi) \cr
&\quad\quad\quad\quad\quad\quad\quad\quad
-\sh (u - \xi) \ch (u - \eta + \xi) \cr} \right) \,, \eqnum $$
where
$$\rho (u, \xi) = \left[ \sh(u + \xi) \sh(-u + \xi) \ch(u - \eta + \xi)
\ch(-u - \eta + \xi) \right]^{-{1\over2}} \,. $$

For other trigonometric $R$ matrices (and in particular, for those
enumerated by Bazhanov and Jimbo), finding the general solution of the
corresponding reflection-factorization equation remains an interesting
open problem. Nevertheless,
because of the relation \eqref{jimboidentity}, the ``trivial'' $K$ matrix
$$ K(u) = 1  \eqlabel{trivial} $$
is a particular solution of the reflection-factorization equation for all the
Bazhanov-Jimbo $R$ matrices except $D^{(2)}_n$, in the homogeneous gauge.
(The $A^{(1)}_1$ $K$ matrices
\eqref{k1}, \eqref{k2} are gauge-equivalent to the identity matrix, in the
limits
$\xi \rightarrow \pm \infty$.) As we shall see in the next section, this
solution is important for constructing models with quantum algebra symmetry.

\vfill\eject

\noindent
\chapnum . INTEGRABLE MODELS WITH QUANTUM ALGEBRA SYMMETRY
\vskip 0.2truein

Having established a generalization of the Zamolodchikov algebra
corresponding to scattering with walls, we turn to the construction
of integrable open quantum spin chains. We shall find a large class
of such models which has quantum algebra symmetry.

As it is well known${}^{\refref{qism}}$, given an arbitrary solution
$R(u)$ of the Yang-Baxter equation, one
can construct an integrable {\it closed} chain, with Hamiltonian
$$H = \sum_{k=1}^{N-1} H_{k,k+1} + H_{N,1} \,, \eqnum  $$
where
$$ H_{k,k+1} = {d\over du} \check R_{k,k+1}(u) \Big\vert_{u=0} \,.
\eqlabel{twosite}  $$
This Hamiltonian, whose state space is $\prod_{k=1}^N \otimes C^n$,
contains only nearest-neighbor interactions. The basic
algebraic structure behind the integrability of this chain is
the intertwining relation
$$R_{12}(u - v)\ \ontop1T(u)\ \ontop2T(v) =
  \ontop2T(v)\ \ontop1T(u)\ R_{12}(u - v) \,, \eqlabel{fundamental} $$
where the monodromy matrix $T(u)$ is given by
$$T(u) = R_{0 N}(u)\ R_{0 N-1}(u)\ \cdots R_{0 1}(u) \,.
\eqlabel{monodromy} $$
The transfer matrix
$$t(u) = \tr T(u) \,, \eqlabel{trace} $$
from which the Hamiltonian is constructed,
plays the role of Cartan generators for the above algebraic structure.
(The trace in \eqref{trace} is over the auxiliary space, which is
denoted by 0 in \eqref{monodromy}.)

In order to construct {\it open} integrable spin chains, a generalization
of the above structure is necessary. The clue to the
introduction of the new transfer matrix is that, while closed chains
are related to scattering on an infinite line ($\sim$ circle), open
chains are related to scattering on an interval. One finds that
the class of $R$ matrices to be used
must now be restricted to those satisfying $PT$-invariance \eqref{pt},
unitarity \eqref{unitarity}, and crossing symmetry \eqref{crossing}. The
Sklyanin transfer matrix is${}^{\refref{sklyanin}, \refref{nonsymmetric}}$
$$t(u) = \tr K_+(u)\ T(u)\ K_-(u)\ T^{-1}(-u) \,, \eqlabel{transfer} $$
where
$$K_-(u) = K(u, \xi_-) \,, \quad\quad\quad
  K_+(u) = K^t (-u -\rho, \xi_+)\ M \,,
\eqnum $$
and $K(u, \xi)$ is a one-parameter ($\xi$) family of solutions of the
the reflection-factorization equation. Indeed, one can show that
$$\left[ t(u) \,, t(v) \right] = 0 \hbox{   for all   } u \,, v
\,. \eqnum $$
Also, $t(u)$ is invariant under gauge transformations \eqref{gauge},
\eqref{gaugek}.
{}From this transfer matrix, one can construct the local Hamiltonian
$$
H = \sum_{k=1}^{N-1} H_{k,k+1}
+ {1\over 2} {d\over du} \ontop1K_-(u) \Big\vert_{u=0}
+ {\tr_0 \ontop0K_+(0) H_{N,0}\over \tr K_+(0)} \,, \eqlabel{openh}
$$
where $H_{k,k+1}$ is given by \eqref{twosite}.

The novel algebraic structure is generated by
$${\cal T}_-(u) = T(u)\ K_-(u)\ T^{-1}(-u) \,, \eqlabel{novel} $$
which, like $K_-(u)$, obeys the reflection-factorization equation
\eqref{sklyanin}. It is not known whether the reflection-factorization
equation admits solutions other than those of the form \eqref{novel}.

Our interest in open quantum spin chains lies in their connection with quantum
algebras. Indeed, let us consider the $R$ matrices of type $g^{(k)}$ which
are listed by Bazhanov and Jimbo. As we have already noted, all of these except
for $A^{(1)}_n$ $(n > 1)$ fulfill the criteria of $PT$ symmetry, unitarity,
and crossing symmetry. Moreover, except for $D^{(2)}_n$, the
reflection-factorization equation
has the trivial solution \eqref{trivial}, which implies
$$K_-(u) = 1 \,, \quad\quad\quad K_+(u) = M \,. \eqlabel{choice} $$
For these choices of $K_\mp$, the Hamiltonian \eqref{openh} reduces to
$$H = \sum_{k=1}^{N-1} {d\over du} \check R_{k,k+1}(u) \Big\vert_{u=0}
\,. \eqlabel{h} $$
In showing that the last term in \eqref{openh} gives only a $c$-number
contribution, one can use the identity
$$\tr_0  \ontop0M \ \check R_{N,0}(u) = f(u)\ \ontop{N}1 \,, \eqnum $$
(where $f(u)$ is a scalar function of $u$) which follows from the
degeneration of \eqref{jimboidentity} at $u = -\rho$ and crossing symmetry.
{}From the identity \eqref{namely}, we conclude that the Hamiltonian commutes
with
the generators of the quantum algebra $U_q[g_0]$,
$$\left[ H \,, U_q[g_0] \right] = 0 \,. \eqnum $$

It is useful to formulate the quantum algebra invariance of these chains in the
$R$ matrix approach. The key${}^{\refref{k/s}}$ is to establish a connection
between the quantum algebra generators and the monodromy matrix $T(u)$. To this
end, we define $R_\pm$ as before \eqref{rplusminus}, and similarly, we set
$$T_\pm = \lim_{u \rightarrow \pm \infty} T(u) \,. \eqnum $$
By taking the limits $u \rightarrow \pm \infty$ and then
$v \rightarrow \pm \infty$ in the fundamental relation \eqref{fundamental},
we obtain the relations
$$
R_\pm \ \ontop1T_\pm \ \ontop2T(v) =
\ontop2T(v)\ \ontop1T_\pm \ R_\pm
\,, \eqlabel{f1}
$$
and
$$
R_\pm \ \ontop1T_\pm \ \ontop2T_\epsilon =
\ontop2T_\epsilon \ \ontop1T_\pm \ R_\pm
\,, \quad\quad \hbox{with} \quad \epsilon = \{+ \,, -\} \,. \eqlabel{f2}
$$
We recognize the latter relation as the definition of the quantum algebra
$U_q[g_0]$. Indeed, the entries of $T_\pm$ can be
expressed${}^{\refref{faddeev}}$ in terms of quantum algebra generators,
as in the $A^{(1)}_1$ example \eqref{borel}. (For the $A^{(2)}_2$ case,
see Ref. \refref{analytic}.)
The relation \eqref{f1} can be interpreted as defining the
tensor character of the operator $T(v)$ with respect to the quantum algebra.
(See also Ref. \refref{tensor}.) We remark that the identity
$$\left[ \check  R_{12}(w) \,, \ontop1T_\pm\ \ontop2T_\pm \right] = 0
\eqlabel{neat}  $$
can also be obtained from the fundamental relation \eqref{fundamental}, if
one instead takes the limits $u,v \rightarrow \pm \infty$ with
$u-v = w = $ finite.

Evidently, in the $R$ matrix approach, quantum algebra symmetry is expressed
through commutators with $T_\pm$. For the open chain transfer matrix
$$t(u) = \tr M\ T(u)\  T^{-1}(-u)  \eqlabel{t} $$
corresponding to the $K$ matrices \eqref{choice}, one can establish
the result${}^{\refref{transfer}}$
$$\left[ t(u) \,,  T_{\pm}  \right] = 0 \,. \eqlabel{result}  $$
This directly implies that
$$\left[ t(u) \,,  U_q[g_0]  \right] = 0 \,. \eqlabel{maybe}   $$
That is, not only the Hamiltonian, but also the transfer matrix
commutes with the quantum algebra generators.

Although we have argued that the transfer matrix \eqref{t} is a
quantum algebra invariant, this invariance is not manifest. Clearly, it
would be desirable to find a ``tensor calculus'' formulation of the
differential geometry on quantum groups, using which the expression for
the transfer matrix would be manifestly invariant. Perhaps this may
be accomplished within the noncommutative geometry of
Connes${}^{\refref{connes}}$ recently
pursued by Wess and Zumino${}^{\refref{wess}}$.
Such a formulation
would undoubtedly lead to the construction of additional invariants.

Having constructed a large class of integrable models with quantum
algebra symmetry, let us briefly comment on their solutions. In the case
of $A^{(1)}_1$, the eigenvalues of the transfer matrix and the Bethe
Ansatz (BA) equations have been determined by the algebraic BA, for
both spin 1/2 ${}^{\refref{sklyanin}}$ and spin 1 ${}^{\refref{spin1}}$.
(The spin 1/2 model was first solved by the coordinate BA
in Ref. \refref{alcaraz}.) Actually, these papers contain the solution
of more general open spin chains, with a two-parameter ($\xi_- , \xi_+$)
class of boundary terms, corresponding to the more general solutions
\eqref{k1}, \eqref{k2} of the reflection-factorization equation.
At least a large class of the general $g^{(k)}$
models with $U_q[g_0]$ symmetry may be solved${}^{\refref{analytic}}$ by
the analytic BA. The BA equations for these open chains
are ``doubled'' with respect to the BA equations for
the corresponding closed chains.

\vskip 0.4truein

\noindent
\chapnum . STRING SOLUTIONS FOR $A^{(2)}_2$ (with A.M. Tsvelik)
\vskip 0.2truein

Determining the Bethe Ansatz (BA) equations and finding the eigenvalues of
the transfer matrix constitute only the first step in calculating
quantities of physical interest for the $g^{(k)}$ models. In order to
investigate the low-temperature thermodynamics of such models, one must
first solve the corresponding BA equations in the
limit that the number of spins $N$ tends to infinity. (See, e.g., Ref.
\refref{tsvelik/wiegmann}.)
Since in the critical regime ($|q|=1$) these models have non-Hermitian
Hamiltonians, one must then make suitable projections on the space of
states.

For the spin 1/2 $A^{(1)}_1$ model, the BA equations in the
$N \rightarrow \infty$ limit have the well-known string solutions
of Takahashi-Suzuki${}^{\refref{takahashi}}$. Moreover, for this model, only
the
boundary terms are non-Hermitian, as is the case in the
Feigin-Fuchs-Dotsenko-Fateev${}^{\refref{dotsenko}}$ construction. While
details
of the projections remain obscure, there is evidence that one
obtains${}^{\refref{alcaraz}, \refref{pasquier}}$
the $c<1$ unitary rational conformal field theories for $q$ a primitive root
of unity. For spin $s \ge 1/2$, the corresponding conformal field theories
are presumably the $SU(2)\otimes SU(2)/SU(2)$
coset models${}^{\refref{rsos}, \refref{cosets}}$.

In the generic case, the picture is less clear. In general, the string
solutions of the BA equations are not known. Also, the bulk terms of the
Hamiltonian (i.e., not just the boundary terms) are non-Hermitian. Such
complications appear already for the case $A^{(2)}_2$.

In an effort to begin to understand these issues, we look here for string
solutions for the $A^{(2)}_2$ model. We expect that
the string solutions for this $U_q[su(2)]$-invariant open chain are a
subset of those for the corresponding closed chain. We therefore focus on
the BA equations of the closed
chain${}^{\refref{vichirko},\refref{tarasov}}$, which are easier
to study:
$$ \eqalignno{
\left[ {\sh \eta \left( \lambda_k + i/2 \right) \over
           \sh \eta \left( \lambda_k - i/2 \right)} \right]^N
& = \prod_{j \ne k}^M {\sh \eta \left( \lambda_k - \lambda_j + i \right) \over
\sh \eta \left( \lambda_k - \lambda_j - i \right)}
{\ch \eta \left( \lambda_k - \lambda_j - i/2 \right) \over
 \ch \eta \left( \lambda_k - \lambda_j + i/2 \right)} \,, \cr
& \quad\quad\quad k = 1\,, \cdots \,, M \,,
  \quad\quad\quad M = 1 \,, 2\,, \cdots \,. \eqalignlabel{ba} \cr} $$

We look for complex solutions
$$\lambda_k = x_k + i y_k \,, \eqnum $$
with $x_k$, $y_k$ real.
We first consider the critical regime $\eta =$ real, with $0 < \eta < \pi$.
(The precise normalization of $\eta$ is not important for our exploratory
discussion.) Clearly, $y_k$ is determined modulo $\pi/\eta$.
Our general strategy is to work with the modulus of these equations. In
particular, we proceed in three steps:
\medskip

\noindent
(1). We fix the value of $M$. Taking the modulus of \eqref{ba}, we obtain
$$\left[ 1 +
{\sin \eta \sin 2\eta y_k \over \sh^2 \eta x_k + \sin^2 \eta (y_k-1/2)}
\right]^N $$
$$\eqalignno{
& = \prod_{j \ne k}^M {\sh^2 \eta (x_k - x_j) + \sin^2 \eta (y_k - y_j + 1)
\over \sh^2 \eta (x_k - x_j) + \sin^2 \eta (y_k - y_j - 1)}\
{\sh^2 \eta (x_k - x_j) + \cos^2 \eta (y_k - y_j - 1/2) \over
\sh^2 \eta (x_k - x_j) + \cos^2 \eta (y_k - y_j + 1/2)} \,, \cr
& \quad\quad\quad  k = 1\,, \cdots \,, M \,. \eqalignlabel{modulus} \cr}$$
Consider the limit $N \rightarrow \infty$. Provided $x_k$, $y_k$ have a finite
limit, the left-hand-side can have one of three possible values:
$$\left\{
\eqalign{& 0 {\hbox{      when      }} \sin \eta \sin 2\eta y_k < 0 \,, \cr
         & 1 {\hbox{      when      }} \sin \eta \sin 2\eta y_k = 0 \,, \cr
         & \infty {\hbox{      when      }} \sin \eta \sin 2\eta y_k > 0 \,.
\cr} \right. $$
Correspondingly, for finite $M$, the right-hand-side must have a zero, be
equal to one, or have a pole. The only way that the right-hand-side can have
a zero or pole for finite values of $x_k$, $y_k$ is for one of the factors
in the numerator or denominator to vanish. This implies certain relations
among $\{x_k \,, y_k\}$. These relations suggest that we look for solutions
of the string type -- that is, sets of solutions
$\{ \lambda_1 \,, \cdots \,, \lambda_m \}$ with a common real part
( the ``center'' ) $x_0$,
$$ \lambda_k = x_0 + i y_k \,, \quad\quad\quad k = 1 \,, \cdots \,, m \,.
\eqnum $$
We shall assume that the center $x_0$ can vary continuously in the
$N \rightarrow \infty$ limit. The task now is to determine $\{ y_k \}$.
\medskip

\noindent
(2). Multiplying together all $M$ BA equations \eqref{ba} together, we obtain
$$\prod_{k=1}^M \left[ {\sh \eta (\lambda_k + i/2) \over
                        \sh \eta (\lambda_k - i/2)} \right]^N = 1 \,, \eqnum $$
which implies
$$\prod_{k=1}^M \Big\vert {\sh \eta (\lambda_k + i/2) \over
                          \sh \eta (\lambda_k - i/2)} \Big\vert^2 = 1 \,.
\eqnum $$
For a single string of length $M$, it follows that
$$\prod_{k=1}^M {\sh^2 \eta x_0 + \sin^2 \eta (y_k + 1/2) \over
\sh^2 \eta x_0 + \sin^2 \eta (y_k - 1/2)} = 1 \,. \eqnum $$
{}From the assumption that $x_0$ is arbitrary, we obtain the set of $M$
relations
$$\sum_{k=1}^M \left(\alpha_k \right)^n =
  \sum_{k=1}^M \left(\beta_k \right)^n \,, \quad\quad\quad n= 1 \,, \cdots \,,
M \,, \eqnum $$
where
$$ \alpha_k = \sin^2 \eta (y_k + 1/2) \,, \quad\quad\quad
   \beta_k = \sin^2 \eta (y_k - 1/2)  \,. $$
The only solutions of the above set of relations are
$$ \left( \alpha_1 \,, \alpha_2 \,, \cdots \,, \alpha_M \right) =
permutation \left( \beta_1 \,, \beta_2 \,, \cdots \,, \beta_M \right) \,.
\eqnum $$
In particular, for $\alpha_k = \beta_j$, it follows that
$$y_k + y_j = { m \pi \over \eta} \quad\quad {\hbox{or}} \quad\quad
  y_k - y_j + 1 = { m' \pi \over \eta} \,, \eqnum $$
with $m$, $m'$ integers.
\medskip

\noindent
(3). Having restricted the possible values of $\{y_k\}$, we return to the
full set of equations \eqref{modulus}, which -- at least for small values
of $M$ -- completely determine $\{ y_k \}$ according to the condition
of their being poles or zeroes of the right-hand-side.

In this way, we have searched for strings of length $M$ for low values of
$M$. Our results are as follows:

\noindent
$M = 1$: There are two strings of length 1, given by
$\lambda = x_0 + iy$ with
$$y = 0 \quad\quad {\hbox{and}} \quad\quad y = {\pi\over 2 \eta} \,, \eqnum $$
respectively.  These are the so-called strings of positive and negative
parity of Takahashi-Suzuki.

\noindent
$M = 2$: There are two strings of length 2, given by
$\lambda_k = x_0 + iy_k$ with
$$y_1 = {i\over 2} \,, \quad\quad\quad y_2 = -{i\over 2} \,, \eqlabel{first} $$
and
$$y_1 = i\left(-{1\over 4} + {\pi\over 4\eta} \right)
\,, \quad\quad\quad
  y_2 = i\left({1\over 4} - {\pi\over 4\eta} \right) \,,  \eqlabel{second} $$
respectively. The first string \eqref{first} is the positive-parity
2-string of Takahashi-Suzuki. The second string, which does not appear
for $A^{(1)}_1$, has been studied numerically in Ref. \refref{nienhuis}.

\noindent
$M = 3$: We find the 3-strings of Takahashi-Suzuki,
$$y_1 = 1 \,, \quad y_2 = 0 \,, \quad y_3 = -1 \quad\quad
(0 < \eta < {\pi\over 2} ) \,,
\eqnum $$
and
$$y_1 = 1 + {\pi \over 2\eta} \,, \quad y_2 = {\pi \over 2\eta} \,,
\quad y_3 = -1 + {\pi \over 2\eta} \quad\quad ({\pi\over 2} < \eta < \pi ) \,,
\eqnum $$
of positive and negative parity, respectively.
We also find the solution
$$y_1 = -{1\over 2} + {\pi \over 2\eta} \,, \quad y_2 = 0 \,,
\quad y_3 = {1\over 2} - {\pi \over 2\eta}  \,, \eqnum $$
which can be interpreted as a combination of a negative-parity 2-string
and a positive-parity 1-string.

\noindent
$M = 4$: We find the positive-parity 4-string of Takahashi-Suzuki,
$$ \{ y_k \} = \{ {3\over 2} \,, {1\over 2} \,, -{1\over 2} \,, -{3\over 2} \}
\quad ( 0 < \eta < {\pi\over 3}  {\hbox{     and      }}
{2\pi\over 3} < \eta < \pi ) \,. \eqnum $$
In addition, we find four new candidate 4-strings:
$$ \{ y_k \} = \{ {5\over 4} - {\pi\over 4 \eta} \,,
{1\over 4} - {\pi\over 4 \eta} \,, -{1\over 4} + {\pi\over 4 \eta} \,,
-{5\over 4} + {\pi\over 4 \eta} \}
\quad\quad ( {\pi\over 5} < \eta < {\pi\over 3}  {\hbox{     and      }}
{\pi\over 2} < \eta < {3\pi\over 5} ) \,, \eqnum $$
$$ \{ y_k \} = \{ 1 - {\pi\over 2 \eta} \,,
{1\over 2}  \,, -{1\over 2} \,, -1 + {\pi\over 2 \eta} \}
\quad\quad ( {\pi\over 3} < \eta < {\pi\over 2} ) \,, \eqnum $$
$$ \{ y_k \} = \{ {3\over 4} + {\pi\over 4 \eta} \,,
{1\over 4} - {\pi\over 4 \eta} \,, -{1\over 4} + {\pi\over 4 \eta} \,,
-{3\over 4} - {\pi\over 4 \eta} \} \,, \eqlabel{q1} $$
$$ \{ y_k \} = \{ {3\over 4} - {\pi\over 4 \eta} \,,
{1\over 4} + {\pi\over 4 \eta} \,, -{1\over 4} - {\pi\over 4 \eta} \,,
-{3\over 4} + {\pi\over 4 \eta} \} \,. \eqlabel{q2} $$
For \eqref{q1}, \eqref{q2}, the analysis of the BA equations is quite
intricate, and we have not been able to confirm that these string
configurations are in fact solutions.

The group-theoretic significance of these new strings has so far eluded us.
The fact that in string configurations
there occur steps of both 1 and ${1\over 2}$, accompanied by necessary
factors of ${\pi\over 4\eta}$, makes it difficult to formulate a
general string hypothesis. This impedes further progress in computing
the thermodynamic properties of this model.

On the other hand, in the noncritical regime $\eta = $ pure imaginary,
the situation is much simpler. Let us make the replacement
$\eta \rightarrow i \eta $ (with $\eta$ real) in the BA equations \eqref{ba}.
Evidently, $x_k$ is determined modulo ${\pi/\eta}$. Repeating
the steps (1) - (3) in the above analysis, we find only the positive-parity
$M$-strings of Takahashi-Suzuki; i.e., $\lambda_k = x_0 + iy_k$ with
$$ \{ y_k \} = \{ {M-1\over 2} \,, {M-3\over 2} \,, \cdots
 {3-M\over 2} \,, {1-M\over 2} \}  \,. \eqnum $$
We do not expect significant difficulties in calculating thermodynamic
properties in this regime.

\vskip 0.4truein

\noindent
\chapnum . CONCLUSIONS
\vskip 0.2truein

We have obtained a number of results concerning integrable spin chains
in connection with quantum algebras. We have presented a generalization of the
Zamolodchikov algebra which accommodates reflecting walls, and which reproduces
the algebraic relations that are obeyed by the $K$ matrices. By either directly
solving these relations or implementing a fusion procedure,
we have obtained new $K$ matrices corresponding to the trigonometric
$R$ matrices for certain (graded) Lie algebras. We have extended Sklyanin's
approach for constructing integrable open quantum spin chains to $PT$-invariant
$R$ matrices, and we have used this formalism to construct and investigate
a large class of models with quantum algebra symmetry. These models may
be solved by the analytic Bethe Ansatz. Finally, we have exhibited new
types of string solutions for the $A^{(2)}_2$ model of Izergin and Korepin.

We are frustrated by the difficulty of solving the Bethe Ansatz equations,
even in the $N \rightarrow \infty$ limit. These equations have a ``group
theoretical'' origin, as they implement the construction of irreducible
representations of a certain algebraic structure. Therefore, there should be
a straightforward algorithm for obtaining their solutions. This, in turn,
should enable one to standardize the calculations of
thermodynamic properties, such as specific heat and magnetic susceptibility,
of the corresponding models.

\vskip 0.4truein

\noindent
\chapnum . ACKNOWLEDGMENTS
\vskip 0.2truein

We are indebted to T. Curtright, P. Freund, M. Jimbo, E. Kiritsis,
P. Kulish, E. Melzer,
N. Reshetikhin, V. Rittenberg, A. Tsvelik and A. Zamolodchikov
for valuable discussions.
This work was supported in part by the National Science Foundation
under Grant PHY-90 07517.

\chapternumberstrue
\chapno=-1

\vskip 0.4truein

\noindent
APPENDIX
\vskip 0.2truein
\bumpchapno

The solution of the graded Yang-Baxter equation corresponding to
$sl(2|1)^{(2)}$ in the fundamental representation is given
by${}^{\refref{kulish/sklyanin}, \refref{bazhanov/shadrikov}}$
$$R(u) = \left( \matrix{a  \cr
                        & b & & r & & & & & y \cr
                        & & c & & & & x       \cr
                        & r & & b & & & & & y \cr
                        & & & & a             \cr
                        & & & & & c & & x     \cr
                        & & x & & & & c       \cr
                        & & & & & x & & c     \cr
                        & y & & y & & & & & d \cr} \right) \,,
\eqlabel{gradedrmatrix}  $$
where
$$\eqalignno{
& a= 1 \,,
\quad\quad b= {\sh u \ \ch(u - \eta)\over \sh(u + 2\eta) \ch(u + \eta)}\,,
\quad\quad c= {\sh u \over \sh(u + 2\eta)} \,, \cr
& d= {1\over \sh(u + 2\eta)} \left[ \sh u -
{\ch \eta \ \sh 2\eta \over \ch (u + \eta)} \right] \,, \quad\quad
  r= {\ch \eta \sh 2\eta \over \sh(u + 2\eta) \ch (u + \eta)} \,, \cr
& x = {\sh  2\eta \over \sh(u + 2\eta)} \,,
\quad\quad\quad\quad y = {\sh u \sh 2 \eta \over \sh(u + 2\eta) \ch( u + \eta)}
\,. \eqalignnum \cr} $$
The parity assignments are given by $p(1) = p(2) = 0 \,, p(3) = 1$.

\vfill\eject

\noindent
8. REFERENCES
\vskip 0.2truein

\reflabel{s}
E.K. Sklyanin, Funct. Anal. Appl. {\bf 16} (1982) 263; {\bf 17} (1983) 273.
\reflabel{kulish/reshetikhin}
P.P. Kulish and N.Yu. Reshetikhin, J. Sov. Math. {\bf 23} (1983) 2435.
\reflabel{drinfeld}
V.G. Drinfel'd, Sov. Math. Dokl. {\bf 32} (1985) 254; {\bf 36} (1988) 212;
J. Sov. Math. {\bf 41} (1988) 898.
\reflabel{jimbo}
M. Jimbo, Commun. Math. Phys. {\bf 102} (1986) 537;
{\it Lecture Notes in Physics}, Vol. 246 (Springer, 1986) 335.
\reflabel{jimboq}
M. Jimbo, Lett. Math. Phys. {\bf 10} (1985) 63; {\bf 11} (1986) 247;
Int'l J. Mod. Phys. {\bf A4} (1989) 3759.
\reflabel{faddeev}
L.D. Faddeev, N. Yu. Reshetikhin and L.A. Takhtajan, Algbr. Anal.
{\bf 1} (1988) 129;  Algebra Analysis {\bf 1} (1989) 178 (in Russian).
\reflabel{woronowicz}
S.L. Woronowicz, Commun. Math. Phys. {\bf 111} (1987) 613;
{\bf 122} (1989) 125.

\reflabel{witten}
E. Witten, Commun. Math. Phys. {\bf 121} (1989) 351;
Nucl. Phys. {\bf B322} (1990) 629; {\bf B330} (1990) 285.

\reflabel{pasquier}
V. Pasquier and H. Saleur, Nucl. Phys. {\bf B330} (1990) 523.

\reflabel{rcft} See, e.g.,
A. Tsuchiya and Y. Kanie, Adv. Stud. in Pure Math. {\bf 16} (1988) 297;
K.-H. Rehren and B. Schroer, Nucl. Phys. {\bf 312} (1989) 715;
G. Felder, J. Fr\"ohlich and G. Keller, Commun. Math. Phys. {\bf 124}
(1989) 417; {\bf 130} (1990) 1;
G. Moore and N. Seiberg, Phys. Lett. {\bf B212} (1988) 451;
Commun. Math. Phys. {\bf 123} (1989) 177;
G. Moore and N. Yu Reshetikhin, Nucl. Phys. {\bf 328} (1989) 557;
L. Alvarez-Gaum\'e, C. Gomez and G. Sierra, Phys. Lett. {\bf B220} (1989) 142;
Nucl. Phys. {\bf B319} (1989) 155; {\bf B330} (1990) 347;
J.-L. Gervais, Commun. Math. Phys. {\bf 130} (1990) 257; Phys. Lett.
{\bf B243} (1990) 85;
L.D. Faddeev, Commun. Math. Phys. {\bf 132} (1990) 131.

\reflabel{pertb} See, e.g.,
F.A. Smirnov, Int'l J. Mod. Phys. {\bf A4} (1989) 4213;
T. Eguchi and S.K. Yang, Phys. Lett. {\bf B224} (1989) 373;
{\bf B235} (1990) 282;
A. LeClair, Phys. Lett. {\bf B230} (1989) 103;
D. Bernard and A. LeClair, Nucl. Phys. {\bf B340} (1990) 721;
C. Ahn, D. Bernard and A. LeClair, Nucl. Phys. {\bf B346} (1990) 409;
N.Yu. Reshetikhin and F.A. Smirnov, Commun. Math. Phys. {\bf 131} (1990) 157;
F.A. Smirnov, Commun. Math. Phys. {\bf 132} (1990) 415.

\reflabel{vladimir}
M.T. Batchelor, L. Mezincescu, R.I. Nepomechie and V. Rittenberg,
J. Phys. {\bf A23} (1990) L141.
\reflabel{spin1}
L. Mezincescu, R.I. Nepomechie and V. Rittenberg, Phys. Lett. {\bf A147}
(1990) 70;
R.I. Nepomechie, in {\it Superstrings and Particle Theory}, ed. by
L. Clavelli and B. Harms (World Scientific, 1990) 319;
L. Mezincescu and R.I. Nepomechie, in {\it Argonne Workshop
on Quantum Groups}, ed. by T. Curtright, D. Fairlie and C. Zachos
(World Scientific, 1991) 206.
\reflabel{unitarity}
L. Mezincescu and R.I. Nepomechie, Phys. Lett. {\bf B246} (1990) 412.
\reflabel{nonsymmetric}
L. Mezincescu and R.I. Nepomechie, J. Phys. {\bf A24} (1991) L17.
\reflabel{twisted}
L. Mezincescu and R.I. Nepomechie, Int'l J. Mod. Phys. {\bf A}, in press.
\reflabel{transfer}
L. Mezincescu and R.I. Nepomechie, Mod. Phys. Lett. {\bf A6} (1991) 2497.
\reflabel{analytic}
L. Mezincescu and R.I. Nepomechie, Miami preprint (1991).

\reflabel{zamolodchikov}
A.B. Zamolodchikov and Al.B. Zamolodchikov, Ann. Phys. {\bf 120} (1979) 253;
A.B. Zamolodchikov, Sov. Sci. Rev. {\bf A2} (1980) 1.

\reflabel{sogo/akutsu/abe}
K. Sogo, Y. Akutsu and T. Abe, Prog. Theor. Phys. {\bf 70} (1983) 730.

\reflabel{kulish/sklyanin}
P.P. Kulish and E.K. Sklyanin, J. Sov. Math. {\bf 19} (1982) 1596.

\reflabel{belavin/drinfeld}
A.A. Belavin and V.G. Drinfel'd, Funct. Anal. Appl. {\bf 16} (1982) 159;
Sov. Sci. Rev. {\bf C4} (1984) 93;
D.A. Leites and V.V. Serganova, Theor. Math. Phys. {\bf 58} (1984) 16.

\reflabel{kac}
V. Kac, {\it Infinite Dimensional Lie Algebras} (Cambridge University Press,
Cambridge, 1985).

\reflabel{bazhanov}
V.V. Bazhanov, Phys. Lett. {\bf 159B} (1985) 321;
Commun. Math. Phys. {\bf 113} (1987) 471.

\reflabel{bazhanov/shadrikov}
V.V. Bazhanov and A.G. Shadrikov, Theor. Math. Phys. {\bf 73} (1987) 1302.

\reflabel{zamo}
A.B. Zamolodchikov, unpublished.
\reflabel{cherednik}
I.V. Cherednik, Theor. Math. Phys. {\bf 61} (1984) 977.
\reflabel{sklyanin}
E.K. Sklyanin, J. Phys. {\bf A21} (1988) 2375.

\reflabel{fateev}
V.A. Fateev and A.B. Zamolodchikov, Sov. J. Nucl. Phys. {\bf 32} (1980) 298.

\reflabel{fusion}
P.P. Kulish and E.K. Sklyanin, {\it Lecture Notes in Physics} {\bf 151}
(Springer, 1982) 61;
P.P. Kulish, N.Yu. Reshetikhin and E.K. Sklyanin, Lett. Math. Phys. {\bf 5}
(1981) 393.

\reflabel{izergin}
A.G. Izergin and V.E. Korepin, Commun. Math. Phys. {\bf 79} (1981) 303.

\reflabel{qism}
R.J. Baxter, {\it Exactly Solved Models in Statistical Mechanics}
(Academic Press, 1982);
L.D. Faddeev and L.A. Takhtajan, Russ. Math. Surv. {\bf 34} (1979) 11;
J. Sov. Math. {\bf 24} (1984) 241;
A.A. Vladimirov, JINR preprint P17-85-742;
L.D. Faddeev, in {\it Les Houches Lectures 1982} ed. by J.B. Zuber and R. Stora
(North-Holland, 1984) 561;
H.J. de Vega, Int'l J. Mod. Phys. {\bf A4} (1989) 2371.

\reflabel{k/s}
P.P. Kulish and E.K. Sklyanin, J. Phys. {\bf A24} (1991) L435; in
{\it Proc. Euler Int. Math. Inst., 1st Semester: Quantum Groups, Autumn 1990},
ed. by P.P. Kulish, in press.

\reflabel{tensor}
L.C. Biedenharn and M. Tarlini, Lett. Math. Phys. {\bf 20} (1990) 271;
V. Rittenberg and M. Scheunert, Bonn preprint (1991).

\reflabel{connes}
A. Connes, Publ. Math. IHES {\bf 62} (1985) 257.

\reflabel{wess}
J. Wess and B. Zumino, CERN preprint (1990).

\reflabel{alcaraz}
F.C. Alcaraz, M.N. Barber, M.T. Batchelor, R.J. Baxter and G.R.W. Quispel,
J. Phys. {\bf A20} (1987) 6397.
See also M. Gaudin, Phys. Rev. {\bf A4} (1971) 386;
{\it La fonction d'onde de Bethe} (Masson, 1983).

\reflabel{tsvelik/wiegmann}
A.M. Tsvelick and P.B. Wiegmann, Adv. in Phys. {\bf 32} (1983) 453.

\reflabel{takahashi}
M. Takahashi and M. Suzuki, Prog. Theor. Phys. {\bf 48} (1972) 2187;
V.E. Korepin, Theor. Math. Phys. {\bf 41} (1979) 953;
K. Hida, Phys. Lett. {\bf A84} (1981) 338;
M. Fowler and X. Zotos, Phys. Rev. {\bf B25} (1982) 5806.

\reflabel{dotsenko}
B.L. Feigin and D.B. Fuchs, unpublished;
Vl.S. Dotsenko and V.A. Fateev, Nucl. Phys. {\bf B240} (1984) 312;
{\bf B251} (1985) 691; C. Thorn, Nucl. Phys. {\bf B248} (1984) 551.

\reflabel{rsos}
E. Date, M. Jimbo, A. Kuniba, T. Miwa and M. Okado, Nucl. Phys. {\bf B290}
[FS20] (1987) 231;
M. Jimbo, A. Kuniba, T. Miwa and M. Okado, Comm. Math. Phys. {\bf 119} (1988)
543;
V.V. Bazhanov and N. Yu. Reshetikhin, Int'l J. Mod. Phys. {\bf A4} (1989) 115.

\reflabel{cosets}
D. Kastor, E. Martinec and Z. Qiu, Phys. Lett. {\bf B200} (1988) 434;
J. Bagger, D. Nemeschansky and S. Yankielowicz, Phys. Rev. Lett. {\bf 60}
(1988) 389; F. Ravanini, Mod. Phys. Lett. {\bf A3} (1988) 271.

\reflabel{vichirko}
V.I. Vichirko and N.Yu. Reshetikhin, Theor. Math. Phys. {\bf 56}
(1983) 805; N.Yu. Reshetikhin, Lett. Math. Phys. {\bf 7} (1983) 205;
N.Yu. Reshetikhin, Sov. Phys. JETP {\bf 57} (1983) 691.

\reflabel{tarasov}
V.O. Tarasov, Theor. Math. Phys. {\bf 76} (1988) 793.

\reflabel{nienhuis}
M.T. Batchelor, B. Nienhuis and S.O. Warnaar, Phys. Rev. Lett. {\bf 62}
(1989) 2425.

\end